\documentclass{desyproc}

\begin{document}

\title{The rise and fall of the fourth quark-lepton generation}
\author{{\slshape M.I. Vysotsky$^1$}\\[1ex]
$^1$ITEP, 117218 Moscow, Russia}

\contribID{xy}

\desyproc{DESY-PROC-2013-03}
\acronym{HQ2013} 

\maketitle

\begin{abstract}

The existence of the fourth quark-lepton generation is not
excluded by the electroweak precision data. However, the recent
results on the 126 GeV higgs boson production and decay do not
allow an extra generation at least as far as the perturbation
theory can be used.

\end{abstract}

\section{Prehistory}

In the course of 1974 November Revolution $J/\psi$ particle was
discovered, and soon it was understood that it consists of $c
\bar{c}$-quarks. In this way the second quark-lepton generation
($\nu_{\mu}$, $\mu$, $s$, $c$) was completed. Two years later
$\tau$-lepton was found, and in 1978 $\Upsilon (b \bar{b})$-meson
was discovered as well. $t$-quark was found only in 1994, however,
already in the 1980s people started to plan finding the particles
of the next, fourth, quark-lepton generation. And the main
question, of course, was: How heavy are $U$, $D$, and $E$?

\section{SLC, LEP}

In the year 1989 $e^+ e^-$ colliders SLC and LEP started to work
at $\sqrt s = M_Z$, and from the determination of $Z$ invisible
width it soon became clear that only three neutrino exist.
According to the final data $\Gamma({\rm invisible})=499 \pm 1.5$
MeV, while according to the theory it equals $166 \cdot 3=498$
MeV, so there is no space for extra neutrinos. However the
possibility of the heavy fourth generation neutrino with the mass
$m_N > M_Z/2$ is not excluded.

\section{Electroweak precision data}

Since the fourth generation quarks and leptons contribute to the
$W$- and $Z$-boson polarization  operators and since these
contributions do not decouple in the limit of heavy new generation
(which is the essence of the electroweak theory and quite opposite
to the case of QED, where, say, the top quark contribution to the
anomalous magnetic moment of muon is suppressed as $(g-2)_{\mu}
\sim(m_{\mu}/m_t)^2$) one can get the constraints on the 4th
generation from the precision measurements of $M_W, m_t$, and
$Z$-boson parameters.

Indeed, in 1998 volume of the Review of Particle Properties Erler
and Langacker wrote: ``An extra generation is excluded at the
99.2\% CL''\cite{1}. The statement of the published in the year
2000 paper \cite{2} is: ``One extra generation is still allowed``.

\begin{figure}[h]
\centerline{\includegraphics[width=.5\textwidth]{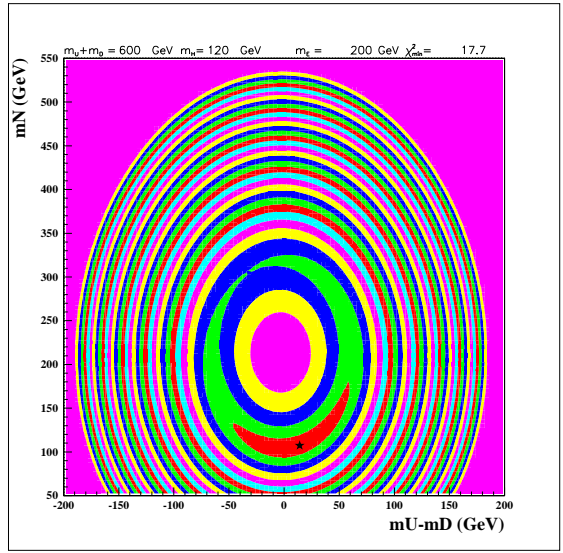}}
\caption{$M_H=120$ GeV, $m_E = 200$ GeV, $m_U + m_D = 600$ GeV,
$\chi^2/d.o.f. = 17.7/11$, the quality of fit is the same as in
SM.} \label{Fig:1}
\end{figure}

\begin{figure}[h]
\centerline{\includegraphics[width=.5\textwidth]{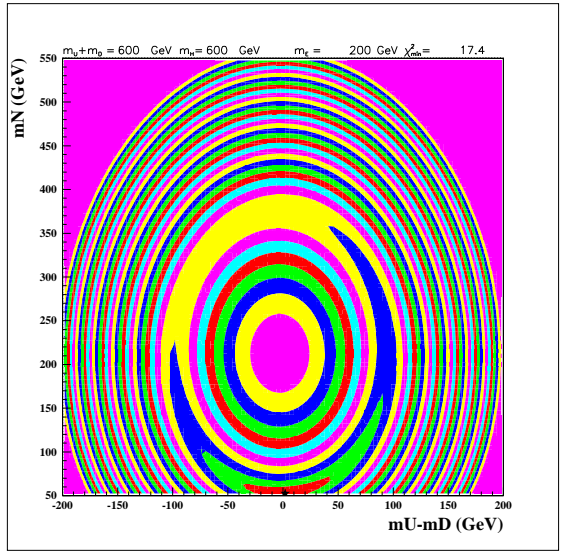}}
\caption{$M_H=600$GeV, $m_E = 200$ GeV, $m_U + m_D = 600$ GeV,
$\chi^2/d.o.f. = 18.4/11$, the quality of fit is the same as in
SM.} \label{Fig:2}
\end{figure}

The following two points were missed by Erler and Langacker:

1. S, T, and U parametrization is valid only when the masses of
all the new particles are much larger than $M_Z$;

2. Instead of making a global fit they studied S, T, and U
separately, while these quantities are correlated. The evolution
of RPP analysis of extra quark-lepton generation in the years 1998
- 2010 is described in detail in paper \cite{3}.

The results of the fit of the electroweak precision observables in
the presence of the fourth generation just before LHC started
obtaining data are shown in Figures 1 and 2 \cite{4}. Fig. 1
corresponds to the light higgs boson, $m_H = 120$ GeV, while Fig.
2 corresponds to heavy higgs, $m_H = 600$ GeV. In both cases the
values of the fourth generation quark and lepton masses are
determined, for which the quality of the fit is practically the
same as for the Standard Model with three generations. In Figures
1 and 2 we put $m_E = 200$ GeV, $m_U + m_D = 600$ GeV, and the
values of $m_N$ and $m_U - m_D$ at which $\chi^2$/d.o.f. is
minimal (and the same as in SM) are shown by star.

\section{LHC direct bounds}

Since the search of heavy quarks  is a relatively easy task for
LHC; the first lower bounds  on their masses appeared soon after
the start of LHC. The last ATLAS bounds are: $m_{t'} > 656$ GeV at
95\% CL if $t' \to Wb$ decay dominates \cite{5} and $m_{b'}
> 480$ GeV if $b' \to Wt$ decay dominates. CMS has similar bounds.
These bounds push heavy quarks out of the perturbative unitarity
domain: $m_{q'}<500$ GeV, so if such quarks exist, their
interaction with the higgs doublet is described by strong dynamics
(let us remind that even for a top quark the coupling with higgs
is not small: $\lambda_t = m_t/(\eta/\sqrt{2}) =
172/(246/\sqrt{2}) \approx 1$.

However, these bounds depend on the pattern of heavy quark decays
and are not universal. Much more interesting indirect bounds
follow from higgs boson production and decay probabilities
measured at LHC.

\section{Higgs data}

In the following Table the values of $\mu$ measured by ATLAS and
CMS collaborations are given. $\mu$ is equal to the ratio of the
measured product of the cross section of $H$ production at LHC and
branching ratio of $H$ decay to a specific final state to the
value of this product calculated in Standard Model. Thus, if there
are no heavy quarks or any other kind of New Physics, $\mu$ equals
one for any decay mode. The data in the Table are taken from
papers \cite{6,7} and correspond to the summer 2013. $H\to bb$
decay was observed only for the associative production of the
higgs boson with $Z$- or $W$-boson.

\bigskip

\centerline{\bf Table}

\bigskip

\centerline{\begin{tabular}{|c|c|c|} \hline  decay mode & ATLAS &
CMS
\\ \hline
$H\to \gamma\gamma$ & $1.6 \pm 0.3$ & $0.77 \pm 0.27$ \\ \hline
$H\to ZZ^*$ & $1.5 \pm 0.4$ & $0.92 \pm 0.28$ \\ \hline  $H\to
WW^*$ & $1.0 \pm 0.3$ & $0.68 \pm 0.20$ \\ \hline $H \to \tau\tau$
& $0.8 \pm 0.7$ & $1.10 \pm 0.41$ \\ \hline $VH\to Vbb$ & $0.2 \pm
0.5$ & $1.00 \pm 0.49$ \\
\hline
\end{tabular}}
\label{tab}

\bigskip

The values of $\mu_i \equiv (\sigma_H \cdot {\rm Br}_i)_{\rm
exp}/(\sigma_H \cdot {\rm Br}_i)_{\rm SM3}$. A new ATLAS result is
$\mu_{\tau\tau} = 1.4 \pm 0.5$.

\bigskip

\begin{figure}[h]
\centerline{\includegraphics[width=.4\textwidth]{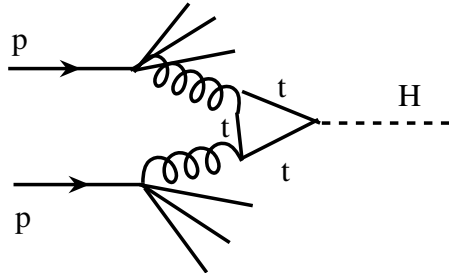}}
\caption{$t\longrightarrow t, t', b'$. $\sigma(gg \longrightarrow
H)_{SM4} \approx 9 \sigma(gg \longrightarrow H)_{SM3}$.}
\label{Fig:3}
\end{figure}

The dominant diagram which describes the higgs boson production at
LHC is shown in Fig. 3. In case of the fourth generation the
amplitude triples since the contributions of heavy $U(t')$ and
$D(b')$ quarks are the same as that of $t$-quark. As a result, the
cross-section of $H$ production in the case of 4 generations is
nine times bigger than in the Standard Model:
\begin{equation}
\sigma(gg \longrightarrow H)_{SM4} \approx 9 \sigma(gg
\longrightarrow H)_{SM3} \label{1}
\end{equation}
Analogously the width of $H\to gg$ decay which in the Standard
Model for $M_H = 126$ GeV is about 0.3 MeV in SM4 becomes 2.7 MeV.
Taking into account that in the Standard Model $\Gamma_H \approx
4.2$ MeV, we get that the branching ratios of $H \to ZZ^*$ and
$H\to WW^*$ decays in the case of SM4 are multiplied by factor
$4.2/6.6 \approx 0.7$, which becomes 0.6 when the modification of
other higgs decay probabilities are taken into account. However,
the electroweak radiative corrections to the $H\to VV$ decay
amplitude being enhanced by factor $(G_F m_{t', b'}^2)$ are big
and according to \cite{8} the factor 0.6 is changed to 0.2 (for
$m_{t'} \approx 600$ GeV) when they are taken into account. It
demonstrates that with such heavy new quarks we leave the domain
of masses generated by Higgs mechanism where the perturbation
theory is applicable. For the value of $\mu$ in case of $H \to
WW^*, ZZ^*$ decays we get the enhancement by factor 2 in the case
of the fourth generation. Such an enhancement is excluded by the
experimental data from the Table. For the lighter fourth
generation quarks the electroweak radiative corrections which
diminish $H \to WW^*, ZZ^*$ decay widths are smaller, so exclusion
will be even stronger.

There is a possibility to diminish $Br(H\to WW^*, ZZ^*)$ by
choosing $M_H/2>m_N>M_Z/2$, which makes $H\to NN$ a dominant higgs
decay mode ($N$ is a neutral lepton of the fourth generation).
From the ATLAS search of $ZH \to l^+l^-$ + invisible decay mode
the 95\% CL upper bound Br($H \to$invisible)$<0.65$ follows
\cite{9}. According to CMS Br($H \to$invisible)$<0.52$. Thus, for
light $N$ the values of $\mu$ for visible final states can be
diminished by factor 2 and for $H\to WW^*$ and $ZZ^*$ decay modes
$\mu$ approaches its SM3 values. Up to now we present the result
of the 4th generation electroweak loop corrections for the
moderate values of the masses of new leptons. If their masses
approach 600 GeV, then factor 0.2 in the suppression of Br($H\to
VV^*$) becomes 0.15 \cite{10} and the value of $\mu$ approaches
its value for the 3 generation case.

\section{$H \to \gamma\gamma$}

In SM3 this decay is described by two one-loop diagrams shown in
Fig. 4. In the limit $M_H << 2 m_t, 2 M_W$ for the decay amplitude
we have:
\begin{equation}
A_3 \sim 7 - 4/3 * 3 * (2/3)^2 = 7 - 16/9 \;\; .  \label{2}
\end{equation}

\begin{figure}[h]
\centerline{\includegraphics[width=.4\textwidth]{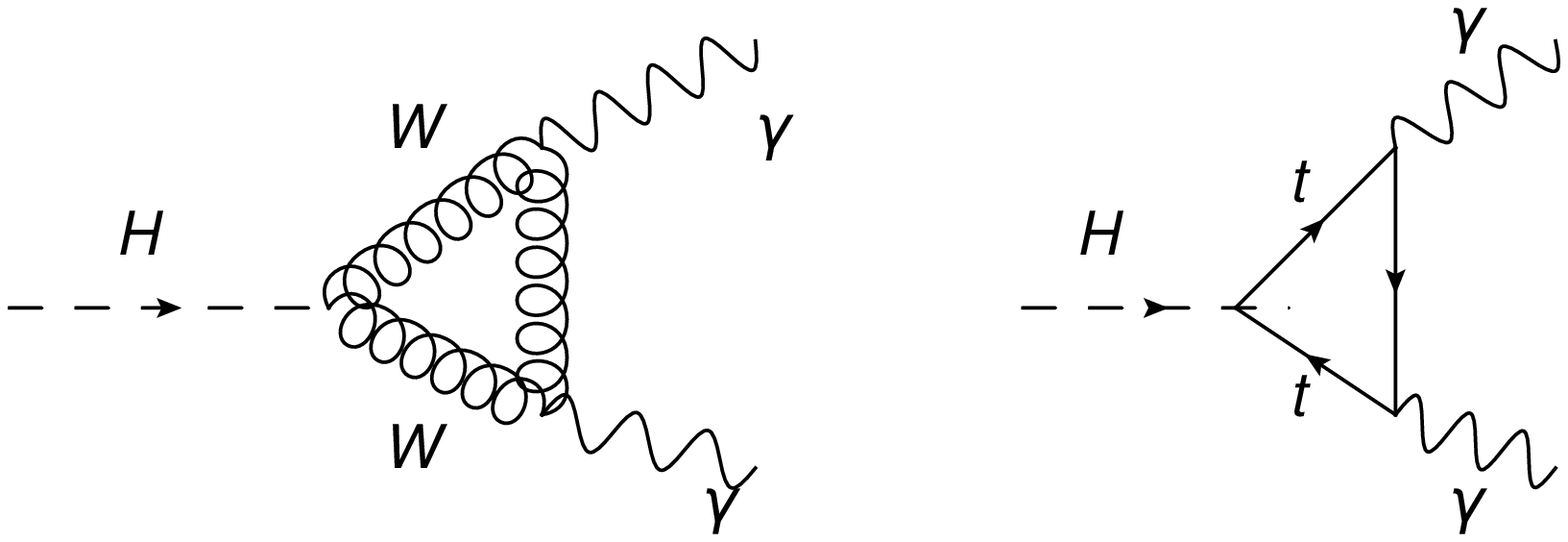}}
\caption{$H\to 2\gamma$ decay in SM3.} \label{Fig:4}
\end{figure}

The numbers 7 and 16/9 are one-loop QED $\beta$ - function
coefficients; the signs correspond to asymptotic freedom and zero
charge behavior, respectively. Number 7 for the first time appears
in 1965 paper of V.S.~Vanyashin and M.V.~Terentiev \cite{11}.
Nowadays it could be derived from the following equation:
\begin{equation}
7 = 22/3 - 1/6 - 1/6 \; , \;\; 22/3 = 11/3 *2 \;\; , \label{3}
\end{equation}
where the factors 1/6 originate from the higgs doublet
contribution into running of SU(2) and U(1) couplings $g$ and
$g'$, while 22/3 is a vector boson contribution into the running
of $g$.

For $M_W = 80.4$ GeV 7 should be substituted by 8.3, while 16/9
has 3\% accuracy for $m_t=172$ GeV. So, in SM3 $A_3 \sim 8.3 -
16/9 = 6.5$, while in the case of the fourth generation a strong
compensation occurs:
\begin{equation}
A_4 \sim 8.3 - 16/9 - 16/9 -4/9 - 4/3 = 3.0 \label{4}
\end{equation}
and taking into account the enhancement of $gg \to H$ production
cross-section and the modification of Higgs decay probabilities
(mainly the enhancement of $H\to gg$ decay), we obtain the same
$\sigma * Br$ ($H\to 2\gamma$) as in SM3:
\begin{equation}
\mu_{2\gamma} = 9 * 0.6 * (3/6.5)^2 \approx 1.2 \;\; . \label{5}
\end{equation}
But the electroweak radiative corrections greatly diminish $\sigma
* {\rm Br} (H \rightarrow 2\gamma)$; according to \cite{8}
it equals $1/3$  of SM3 result or even less, while the average of
ATLAS and CMS data is $1.2\pm 0.2$, so the 4th generation is
excluded at 4-5 $\sigma$ level. It would be good to calculate 3
loop electroweak corrections to the $\Gamma (H\to 2\gamma)$ in the
case of the fourth generation.

\section{$H\rightarrow \tau\tau$, $H \to bb$}

$\mu$ for the ($\tau\tau$) mode at tree level equals approximately
\begin{equation}
\mu_{\tau\tau} \approx 9 \; ({\rm from}\; H \; {\rm production ~
cross ~ section}) * 0.6 \; ({\rm enchancement ~ of} \; H \; {\rm
width ~ in ~ SM4}) \approx 5 \;\; , \label{6}
\end{equation}
and the electroweak loop corrections make the decay width larger
by 30\% \cite{8}. The experimental data exclude this huge
enhancement.

The consideration differs for $H\to b\bar b$ mode: it is seen only
in the associative higgs boson production $VH \to Vbb$, which
unlike gluon fusion is not enhanced in the 4th generation case,
and there is no contradiction with the LHC experimental data.

\section{Conclusions}

\begin{itemize}
\item LHC data on 126 GeV higgs boson production and decays
exclude the Standard Model with the sequential fourth generation
in the perturbative domain: too small $gg\to H \to \gamma\gamma$
probabilitiy, too big $gg\to H \to \tau\tau$ probability;

\item If we are out of the perturbative domain ($m4 \sim 1$ TeV)
extra generation cannot be excluded, but we are unable to
understand why all the experimentally measured $\mu$'s are close
to one and SM3 works so well;

\item In two higgs doublets model the fourth generation is still
allowed \cite{12};

\item Since the vector generation has SU(2)$\times$U(1) invariant
masses it is not excluded by higgs data.

\end{itemize}

\section{Acknowledgments}

I am grateful to Helmholz School organizers for hospitality at
Dubna and to A.N.~Rozanov and I.I.~Tsukerman for useful
discussions and comments. I am supported in part by the RFBR grants
№ 11-02-00441 and 12-02-00193.

\end{document}